\documentclass[twocolumn,showpacs,prl,superscriptaddress]{revtex4}

\usepackage{graphicx}

\newcommand{\figurewidth}{0.48\textwidth}

\begin{document}

\title{Structure and stability of self-assembled actin--lysozyme complexes
in salty water}
\author{Lori K. Sanders}
\affiliation{Dept.\ of Materials Science and Engineering, University of
Illinois at Urbana-Champaign, Urbana, Illinois 61801}
\author{Camilo Gu\'aqueta}
\affiliation{Dept.\ of Materials Science and Engineering, University of
Illinois at Urbana-Champaign, Urbana, Illinois 61801}
\author{Thomas E. Angelini}
\affiliation{Dept.\ of Physics, University of Illinois at
Urbana-Champaign, Urbana, Illinois 61801}
\author{Jae-Wook~Lee}
\affiliation{Dept.\ of Materials Science and Engineering, University of
Illinois at Urbana-Champaign, Urbana, Illinois 61801}
\author{Scott C. Slimmer}
\affiliation{Dept.\ of Materials Science and Engineering, University of
Illinois at Urbana-Champaign, Urbana, Illinois 61801}
\author{Erik Luijten}
\email[Corresponding author. E-mail: ]{luijten@uiuc.edu}
\affiliation{Dept.\ of Materials Science and Engineering, University of
Illinois at Urbana-Champaign, Urbana, Illinois 61801}
\author{Gerard C. L. Wong}
\email[Corresponding author. E-mail: ]{gclwong@uiuc.edu}
\affiliation{Dept.\ of Materials Science and Engineering, University of
Illinois at Urbana-Champaign, Urbana, Illinois 61801}
\affiliation{Dept.\ of Physics, University of Illinois at
Urbana-Champaign, Urbana, Illinois 61801}
\date{\today }

\begin{abstract}
  Interactions between actin, an anionic polyelectrolyte, and lysozyme,
  a cationic globular protein, have been examined using a combination of
  synchrotron small-angle x-ray scattering and molecular dynamics
  simulations.  Lysozyme initially bridges pairs of actin filaments,
  which relax into hexagonally-coordinated columnar complexes comprised
  of actin held together by incommensurate one-dimensional close-packed
  arrays of lysozyme macroions. These complexes are found to be stable
  even in the presence of significant concentrations of monovalent salt,
  which is quantitatively explained from a redistribution of salt
  between the condensed and the aqueous phases.
\end{abstract}

\maketitle

In the presence of multivalent cations, anionic biological
polyelectrolytes can overcome their electrostatic repulsion and exhibit
a mutual attraction.  These ``like-charge attractions'' result from ion
correlations that cannot be understood within mean-field theories such
as the commonly-employed Poisson--Boltzmann formalism
\cite{gelbart00,grosberg02,levin02}. The problem becomes more complex
when the mediating multivalent cations are themselves macroions.
Macroion--polyelectrolyte complexes occur in many physical systems, such
as DNA--dendrimer complexes for nonviral gene therapy~\cite{evans02} and
antimicrobial binding in cystic fibrosis~\cite{weiner03}. Various
factors affect their formation: The presence of salt can lead to an
attraction driven by osmotic pressure~\cite{parsegian95}.  Differential
screening of positive and negative charges distributed on the surface of
a macroion may significantly modify interactions at the
macroion--polyelectrolyte interface~\cite{seyrek03}. Entropic gain due
to mutual neutralization and consequent counterion release upon
macroion--polyelectrolyte ``adhesion'' is expected to be important, but
can be potentially modulated by the steric commensurability between the
charge pattern on the polyelectrolyte and the macroion
size~\cite{henle04}.  The relative importance of all these interactions,
and how they modify one another in their combined effect on the
structural evolution of macroion--polyelectrolyte complexes, is
generally unknown.

\begin{figure}[tbp]
\centerline{\includegraphics[width=\figurewidth]{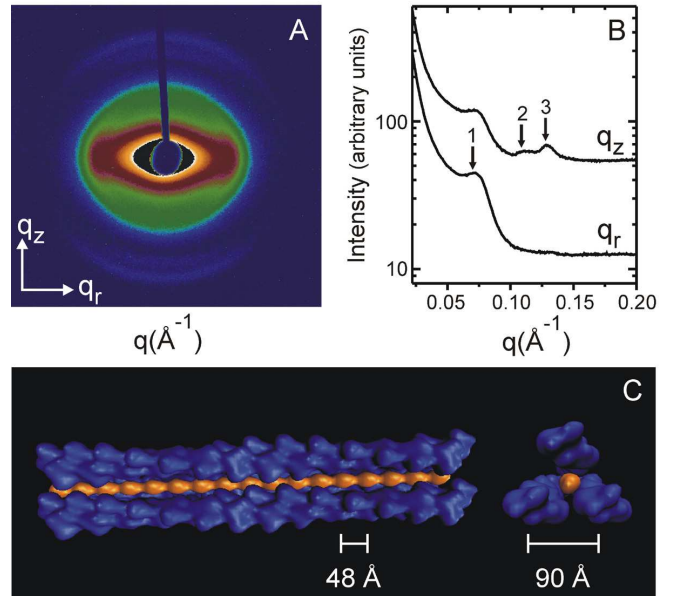}}
\caption{(A) Synchrotron 2D x-ray diffraction pattern of partially aligned
actin--lysozyme bundles, formed in a solution containing 150mM KCl. (B) 1D
integrated slices along the $q_z$ and $q_r$ directions with arrows marking
the actin--actin close-packed bundling peak (1), the actin helix form factor
(2), and the lysozyme--lysozyme correlation peak (3). (C) Proposed structure
of actin--lysozyme composite bundles (side and end views): Lysozyme (orange)
is close-packed in three-fold symmetric sites between actin filaments
(blue). }
\label{fig:saxs}
\end{figure}

In this Letter, we examine the role of several of the above-mentioned
interactions in the complexation of actin and lysozyme, a prototypical
system of oppositely-charged ``rods'' and ``spheres,'' over a range of
monovalent salt concentrations.  Using synchrotron small-angle x-ray
scattering (SAXS), we show that self-assembled complexes are comprised
of hexagonally-coordinated columnar arrangements of actin held together
by \emph{one-dimensional}~(1D) arrays of lysozyme macroions at the
three-fold interstitial ``tunnels'' of the columnar actin sublattice
(Fig.~\ref{fig:saxs}). Molecular dynamics (MD) simulations using a
realistic model of the actin helix provide a detailed confirmation of
this picture, and reveal structural reconstructions and corresponding
salt redistribution within an actin--lysozyme bundle as the inter-actin
separation is varied.  Both experiment and simulation show that the
lysozyme is arranged in a close-packed manner, \emph{incommensurate}
with the actin periodicity. Moreover, the self-assembly of columnar
actin--lysozyme complexes is \emph{enhanced} for higher concentrations
of monovalent ions. We believe that these results can be explained by
significant repartitioning of salt between the condensed and the aqueous
phases, which strongly modifies screening effects.

F-actin is an anionic rod-like cytoskeletal polymer (diameter
7.5\textrm{nm}, charge density $-e/0.25\text{nm}$, persistence length
10$\mu $\textrm{m}).  Lysozyme is approximately an ellipsoid of size
$2.6\text{nm}\times 2.6\text{nm}\times 4.5\text{nm}$ with a net charge
of +$9e$ at neutral pH\@.  Monomeric G-actin (MW 42,000) was prepared
from a lyophilized powder of rabbit skeletal muscle. The
non-polymerizing G-actin solution contained a 5mM TRIS buffer at pH 8.0,
with 0.2mM CaCl$_{2}$, 0.5mM ATP, and 0.2mM DTT and 0.01\% NaN$_{3}$.
G-actin (2mg/ml) was polymerized into F-actin upon the addition of salt
(100mM KCl). Human plasma gelsolin was used to control the average
F-actin length to~$\sim$1mm. The F-actin filaments were treated with
phalloidin (MW 789.2) to prevent depolymerization.  Hen egg white
lysozyme (MW 14,300) was mixed with F-actin in 1.5mm diameter quartz
capillaries to form isoelectric actin--lysozyme complexes.  SAXS
experiments were performed both at Beamline 4-2 of the Stanford
Synchrotron Radiation Laboratory as well as at an in-house x-ray source.
The incident synchrotron x-rays from the 8-pole Wiggler were
monochromatized to 8.98KeV ($\lambda $=1.3806\AA) using a double-bounce
Si(111) crystal, focused using a cylindrical mirror. The scattered
radiation was collected using an MAR Research charge-coupled device
camera (pixel size 79$\mu $m). For the in-house experiments, incident
CuK$_{\alpha }$ radiation ($\lambda $=1.54\AA) from a Rigaku
rotating-anode generator was monochromatized and focused using Osmic
confocal multilayer optics, and scattered radiation was collected on a
Bruker 2D wire detector (pixel size 105$\mu $m). The 2D SAXS data from
both systems are mutually consistent.

A 2D diffraction pattern for partially aligned isoelectric
F-actin--lysozyme bundles and its associated 1D integrated slices along
the $q_{z}$ and $q_{r}$ directions are shown in Figs.\ \ref{fig:saxs}A
and~\ref{fig:saxs}B\@.  Examination of the slice along the equatorial
($q_{r}$) direction shows a correlation peak at $q$=0.07\AA$^{-1}$ that
corresponds to close-packed composite actin--lysozyme bundles. The
inter-actin spacing of 90\AA{} is consistent with a columnar actin
lattice expanded by lysozyme in three-fold interstitial ``tunnels,''
aligned with its long axis parallel to the actin. This inter-actin
spacing is significantly larger than the 75\AA{} spacing for
close-packed actin condensed with multivalent ions~\cite{angelini03}. No
other arrangement of lysozyme and actin will reproduce this diffraction
pattern, given their respective sizes. Along the meridional ($q_{z}$)
direction, a weak, mosaic-smeared actin form factor feature at
0.113\AA$^{-1}$ is observed, as well as a new, strong correlation peak
that differs from expected actin form factor features (layer lines) in
position, orientation, and relative intensity.  The appearance of this
peak at $q_{z}$=0.130\AA$^{-1}$ corresponds to an inter-lysozyme
distance of 48.3\AA, comparable to the length of lysozyme along its
major axis, which suggests that lysozyme is close-packed along this
direction within the bundles.  Interestingly, this lysozyme periodicity
is incommensurate with the projected actin periodicity ($\sim$56\AA), in
contrast with the behavior of divalent ions on actin~\cite{angelini03}.
This incommensurate arrangement permits charge matching between actin
and lysozyme within the bundle, and indicates the important role of
entropy gain from counterion release in this system.
Figure~\ref{fig:saxs}C shows schematic representations of a condensed
bundle.

\begin{figure}[b]
\centerline{\includegraphics[width=\figurewidth]{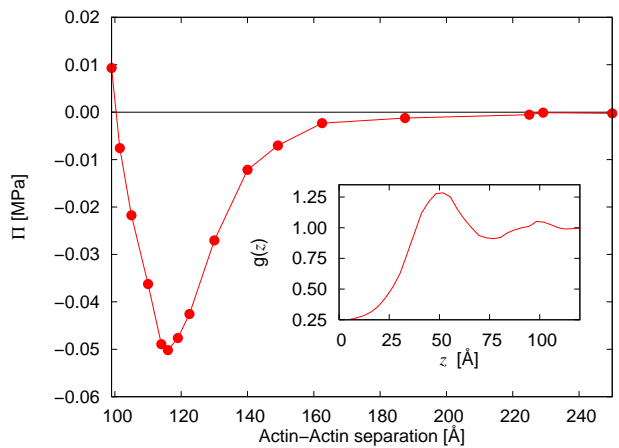}}
\caption{Osmotic pressure of a hexagonally-coordinated bundle of actin
filaments in excess solution, as determined from MD simulations. Inset:
lysozyme pair correlation function along the filament axis. For
discussion see the text.}
\label{fig:osmotic}
\end{figure}

In order to elucidate the underlying mechanism for bundle formation and
the structure of the resulting complex, we have performed MD simulations
using a modified version of Moldy~\cite{moldy}. In these simulations,
G-actin is modeled using the four-sphere model~\cite{al-khayat95}, which
is based upon crystallographic measurements and provides a relatively
accurate coarse-grained representation of the monomer charge
distribution. F-actin is comprised of a sequence of these monomers, in
which successive units have a separation of 27.5\AA{} and a relative
rotation of 166.7$^{\circ}$ around the filament axis. This leads to a
helical structure with a repeat unit of 13 monomers. The filaments are
assembled into a parallel hexagonally-coordinated bundle.  An elementary
simulation cell consists of a bundle fragment containing 2$\times $2
filaments with a length of 6 repeat units (78 monomers) each. This cell
is periodically replicated in all directions. Following the experiments,
we set the lysozyme concentration in the bundle to neutralizing
conditions, which corresponds to 352 lysozyme units per simulation cell.
Each lysozyme is modeled as a rigid dumbbell structure of two spheres
with diameter~$25$\AA{} and charge~$4.5e$, at a center-to-center
distance of $20$\AA, thus approximating the aforementioned ellipsoidal
dimensions.  Additional salt is modeled as monovalent spherical
particles with a hydrated radius of 3.3\AA.  Coulomb interactions are
treated by means of Ewald summation, and excluded-volume interactions
are represented by pair potentials of the form $k_{\rm
B}T(\sigma/r)^{12}$, where $\sigma $ is the sum of the effective radii
of two interacting particles (ions, G-actin subunits and lysozyme
subunits).  During each simulation, the actin separation is fixed,
whereas lysozyme and all ions move freely. Thus, the simulations probe
the stability of a swelling bundle while maintaining the filaments in a
parallel configuration, ignoring their rotational degrees of freedom.
This is justified by the observation that the calculations are confined
to actin separations below 25nm, i.e., less than 0.25\% of the
persistence length. Mutual sliding and axial rotation of the filaments
are not taken into account.

This model indeed predicts electrostatically driven complex formation.
Since the water is modeled as a dielectric continuum, the osmotic
pressure~$\Pi $ can be obtained directly from the virial involving all
interparticle forces~\cite{hill56}. Bundle formation takes place in
excess solution, and hence the bundle stability follows from a
comparison of the osmotic pressure to the osmotic pressure of the
salt~$\Pi_{\mathrm{salt}}$. A negative osmotic pressure difference
$\Delta \Pi \equiv \Pi -\Pi_{\mathrm{salt}}$ implies bundle contraction
and the free-energy minimum [$\Delta \Pi =0$ and $\partial (\Delta \Pi
)/\partial V<0$] yields the stable actin separation. A comparable
approach has been employed before (see
Refs.~\cite{lyubartsev98,deserno03} and references therein) to study the
condensation of rod-like polyelectrolytes by counterions.
Figure~\ref{fig:osmotic} shows that, under salt-free conditions, an
inter-actin spacing of $\sim$100\AA{} is predicted, in quite close
agreement with the experimental observations.

\begin{figure}
\centerline{\includegraphics[width=\figurewidth]{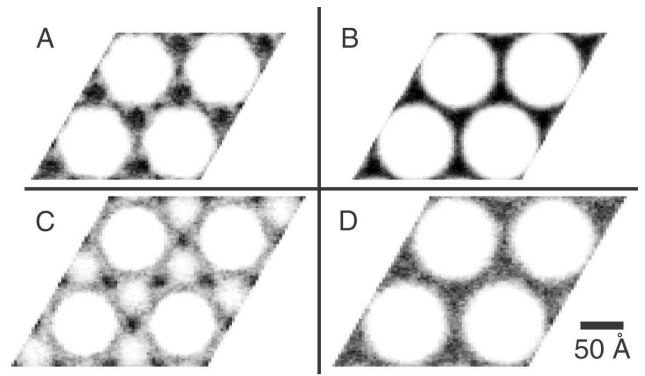}}
\caption{Contour plots showing the lysozyme distribution in actin--lysozyme
complexes without added salt. Darker shading corresponds to higher
concentrations. In the equilibrium configuration (A), lysozyme is
predominantly located in the three-fold interstitial regions. The
free-energy inflection point occurs at a slightly expanded lattice (C), in
which lysozyme is depleted from these regions and occupies the bridging
regions instead. Panels (B) and~(D) are the counterparts of panels (A)
and~(C), respectively, in the absence of electrostatic forces.}
\label{fig:density}
\end{figure}

Having established that our model captures essential aspects of
actin--lysozyme complexation, we exploit it to elucidate the
\emph{structural} properties and evolution of the resulting complexes.
Figure~\ref{fig:density} shows contour plots of the lysozyme center of
mass, projected on a plane perpendicular to the bundle axis. In order to
minimize artifacts resulting from the finite actin length, the
calculations employed filaments consisting of 12 repeat units (156
monomers). In the equilibrium configuration (Fig.~\ref{fig:density}A),
the maximum lysozyme concentration occurs in the three-fold interstitial
regions between the actin filaments, supporting the interpretation of
the SAXS diffraction data in Fig.~\ref{fig:saxs}.  Furthermore, the
lysozyme pair correlation function (inset of Fig.~\ref{fig:osmotic})
along the actin axis shows a clear peak at $z=50$\AA, corroborating the
experimentally measured close-packed value of
$48.3$\AA~\cite{lysozyme-gz}.  Minor enhancements of the lysozyme
concentration can also be seen in the bridging regions between pairs of
neighboring filaments.  The lysozyme distribution is governed both by
entropic effects and by a competition between electrostatic protein
repulsions and actin--lysozyme attractions. Also, the helical actin
structure imposes an excluded-volume repulsion with a rather pronounced
shape. In order to distinguish any structure resulting from this
repulsion, Fig.~\ref{fig:density}B shows a contour plot obtained in a
simulation without electrostatic effects. The homogeneous lysozyme
distribution confirms that the structure in Fig.~\ref{fig:density}A is
dominated by electrostatic interactions.  Interestingly, the actin
excluded volume (white circular regions) also has a more pronounced
hexagonal structure in panel~A, which is caused by the precessing
highly-charged regions on the monomers.  Consideration of larger lattice
spacings provides information on intermediate states that may arise
during the complexation process. As illustrated in
Fig.~\ref{fig:density}C, significant rearrangements occur in the final
stages of the bundle formation: At the osmotic-pressure minimum
(free-energy inflection point), i.e., at an actin separation that is
increased by merely $16$\AA, lysozyme is \emph{depleted} from the
interstitial regions and instead predominantly occupies the bridging
sites.  Figure~\ref{fig:density}D confirms that this lysozyme
distribution is again dominated by electrostatic effects.  Thus, the
complex evolves from lysozyme at bridging positions between pairs of
actin rods to the final three-fold positions
observed by experiment.  Lysozyme and actin ``contact'' interactions are
maintained during this structural relaxation, indicating that their
mutual electrostatic attraction plays an important role.

\begin{figure}
\centerline{\includegraphics[width=\figurewidth]{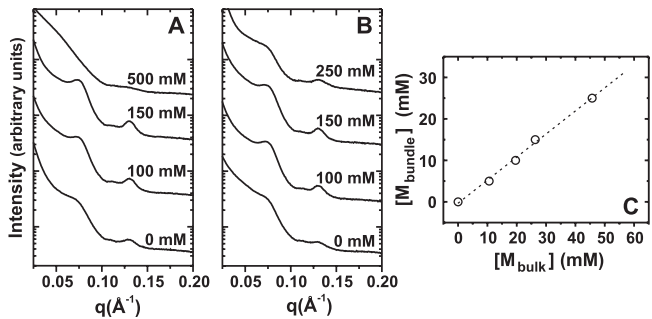}}
\caption{Series of diffraction data showing the evolution of bundle
structure as a function of (A) NaCl and (B) KCl concentration with
maximum bundling occurring around 120mM. (C) Simulated ($\circ$) bundle
vs.\ bulk ion concentrations for a stabilized actin--lysozyme complex.
Salt repartitions into different concentrations inside and outside the
bundle. The dashed line is a guide to the eye.}
\label{fig:diffraction}
\end{figure}

It is interesting to consider how this self-assembly is affected by the
addition of monovalent salt. The counter\-ion release mechanism implicit
in actin--lysozyme binding will strongly modify qualitative arguments
based on screening. The repartitioning of salt between the condensed and
aqueous phases implies a different degree of screening inside and
outside the complex. In addition, it can lead to a stabilizing external
osmotic pressure. Such a redistribution of ions is often not taken into
account theoretically. To study these effects, a series of SAXS
measurements were performed on actin--lysozyme complexes at different
NaCl concentrations. As [NaCl] is increased to 150mM, the turbidity
increases and the intensity of the lysozyme--actin diffraction peak
increases without significant changes in its peak width
(Fig.~\ref{fig:diffraction}A), indicating the formation of more bundles
(rather than bundles that are more ordered and have larger coherent
domains). At higher salt concentrations, the trend reverses and a
weakening of the bundling peak is observed. The same results are found
using KCl (Fig.~\ref{fig:diffraction}B), showing that this is not a
cation-specific binding effect.

Since regular screening is likely to play a role in the ultimate
disappearance of the bundle, we concentrate on the stability at low and
intermediate salt levels.  Simulations of bundles with additional salt
show that the osmotic pressure within the complex rises more rapidly as
a function of salt concentration than the bulk pressure of salt at the
same concentration, leading to destabilization once the salt
concentration exceeds $\sim$10mM\@. This apparent discrepancy with the
experimental findings already suggests a redistribution of salt ions. To
quantify this further, grand-canonical simulations of the bundle can be
employed~\cite{lyubartsev98}. Here, we have chosen an alternative
strategy. Using the Widom particle-insertion
technique~\cite{frenkel-smit2}, we determine the chemical potential of
salt within the complex, as a function of concentration and actin
separation. Exploiting the coexistence condition, we subsequently
determine the corresponding \emph{bulk} salt concentrations and osmotic
pressures via independent grand-canonical Monte Carlo simulations. This
yields several important results.  Firstly, the salt concentration
inside the actin--lysozyme complex is approximately twice lower than the
bulk concentration, which leads to a difference in osmotic pressure that
is sufficient to maintain bundle stability up to much higher
concentrations than otherwise would have been possible
(Fig.~\ref{fig:diffraction}C\@).  Secondly, the depression of the ion
concentration within the bundle may explain why maximal actin--lysozyme
bundling is observed at a \emph{global} salt concentration around 120mM,
rather than at concentrations in the range 30--60mM as predicted from
differential screening arguments~\cite{seyrek03}.


In summary, by studying self-assembled actin--lysozyme complexes via a
combination of small-angle x-ray scattering and molecular dynamics
simulations, we have shown that salt repartitioning impinges strongly on
the structure and stability of the complex, and qualifies
commonly-invoked mechanisms such as counterion release and differential
screening.

\begin{acknowledgments}
  We gratefully acknowledge discussions with M. Olvera de la Cruz, P.
  Dubin, M. Rubinstein, and D. Harries. This material is based upon work
  supported by the U.S. Department of Energy, Division of Materials
  Sciences under Award No.\ DEFG02-91ER45439, through the Frederick
  Seitz Materials Research Laboratory at the University of Illinois at
  Urbana-Champaign and by the National Science Foundation under Grants
  DMR-0346914 (to EL), DMR-0409369 (to GW), and CTS-0120978.
\end{acknowledgments}


\begin{thebibliography}{10}
\bibitem{lysozyme-gz}
To avoid cross-correlations between different three-fold interstitial regions,
  the pair correlation function is computed for each region separately.
\end{thebibliography}

\end{document}